\PassOptionsToPackage{cmyk,table}{xcolor}
\documentclass[conference,pbalance]{iaria} 
\pdfoutput=1 
\DeclareBibliographyCategory{cited}
\AtEveryCitekey{\addtocategory{cited}{\thefield{entrykey}}}

\usepackage[babel=true,english=american]{csquotes}
\usepackage[english]{babel} 
\usepackage[shortcuts]{extdash}

\usepackage[alwaysadjust]{paralist}
\usepackage[caption=false,font=footnotesize]{subfig}
\usepackage{caption}
\usepackage[
  babel=true,
  expansion=alltext,
  protrusion=alltext-nott,
  nopatch=eqnum,
  final
]{microtype}

\usepackage{graphicx}
\usepackage{amsmath}
\usepackage{booktabs}
\usepackage[most]{tcolorbox}
\usepackage{orcidlink} 

\newtcolorbox{keytakeaway}[1][]{
  colback=black!4, colframe=black!55, boxrule=0.5pt, arc=2pt,
  left=5pt, right=5pt, top=4pt, bottom=4pt, boxsep=2pt,
  fonttitle=\bfseries\footnotesize, fontupper=\footnotesize,
  title={#1}, breakable
}
\usepackage{fontawesome} 
\usepackage{relsize}     
\title{CodePori: Large-Scale System for Autonomous Software Development Using Multi-Agent Technology}

\author{%
\begin{tabular}{c@{\hspace{0.012\textwidth}}c@{\hspace{0.012\textwidth}}c@{\hspace{0.012\textwidth}}c}

\begin{tabular}[t]{@{}c@{}}
\normalsize Zeeshan Rasheed\\[-1pt]
\scriptsize Faculty of Information Technology\\[-1pt]
\scriptsize and Communication Sciences\\[-1pt]
\scriptsize Tampere University\\[-1pt]
\scriptsize e-mail: zeeshan.rasheed@tuni.fi
\end{tabular}
&
\begin{tabular}[t]{@{}c@{}}
\normalsize Muhammad Waseem\\[-1pt]
\scriptsize Faculty of Information Technology\\[-1pt]
\scriptsize and Communication Sciences\\[-1pt]
\scriptsize Tampere University\\[-1pt]
\scriptsize e-mail: muhammad.waseem@tuni.fi
\end{tabular}
&
\begin{tabular}[t]{@{}c@{}}
\normalsize Kai-Kristian Kemell\\[-1pt]
\scriptsize Faculty of Information Technology\\[-1pt]
\scriptsize and Communication Sciences\\[-1pt]
\scriptsize Tampere University\\[-1pt]
\scriptsize e-mail: kai-kristian.kemell@tuni.fi
\end{tabular}
&
\begin{tabular}[t]{@{}c@{}}
\normalsize Aakash Ahmad\\[-1pt]
\scriptsize School of Computing\\[-1pt]
\scriptsize University of Derby, UK\\[-1pt]
\scriptsize e-mail: a.abbasi@derby.ac.uk
\end{tabular}
\\

\multicolumn{4}{c}{\rule{0pt}{14pt}}\\[-1pt]

\begin{tabular}[t]{@{}c@{}}
\normalsize Malik Abdul Sami\\[-1pt]
\scriptsize Faculty of Information Technology\\[-1pt]
\scriptsize and Communication Sciences\\[-1pt]
\scriptsize Tampere University\\[-1pt]
\scriptsize e-mail: malik.sami@tuni.fi
\end{tabular}
&
\begin{tabular}[t]{@{}c@{}}
\normalsize Mika Saari\\[-1pt]
\scriptsize Faculty of Information Technology\\[-1pt]
\scriptsize and Communication Sciences\\[-1pt]
\scriptsize Tampere University\\[-1pt]
\scriptsize e-mail: mika.saari@tuni.fi
\end{tabular}
&
\begin{tabular}[t]{@{}c@{}}
\normalsize Jussi Rasku\\[-1pt]
\scriptsize Faculty of Information Technology\\[-1pt]
\scriptsize and Communication Sciences\\[-1pt]
\scriptsize Tampere University\\[-1pt]
\scriptsize e-mail: jussi.rasku@tuni.fi
\end{tabular}
&
\begin{tabular}[t]{@{}c@{}}
\normalsize Pekka Abrahamsson\\[-1pt]
\scriptsize Faculty of Information Technology\\[-1pt]
\scriptsize and Communication Sciences\\[-1pt]
\scriptsize Tampere University\\[-1pt]
\scriptsize e-mail: pekka.abrahamsson@tuni.fi
\end{tabular}

\end{tabular}%
}

\begin{document}

\maketitle

\begin{abstract}
Existing Large Language Model (LLM)-based multi-agent systems are capable of executing tasks and providing data-driven recommendations, thereby enabling automation and decision support that assist practitioners in software development. However, existing studies have evaluated agents’ performance on benchmark datasets, offering only binary pass-or-fail results, which provide limited insight into their practical applicability. There remains a lack of empirical research examining the potential and limitations of LLM-based agents in addressing challenging, real-world tasks, such as automated code generation for software systems. To this end, this study conducted a survey to empirically investigate the potential of LLM-based agents in software development, with participants evaluating the agents’ performance in autonomous software development tasks.
We employed a two-phase approach comprising (i) the development of a multi-agent system, CodePori, to automate code generation, and (ii) a survey-based evaluation to assess agent performance and explore its practical applicability to software development. Our results highlight that, while LLM-based multi-agent systems show potential for autonomous code generation, their successful integration requires addressing specific challenges (e.g., short-term memory limitations, hallucinations, and code smells) and incorporating a practitioner-centric perspective. The study also highlights the need to move beyond standard benchmarks for evaluating real-world applicability and identifies new opportunities for broader adoption in both industry and academia.

\end{abstract}


\section{Introduction}\label{sec:intro}

Software Engineering (SE) as a discipline supports the application of structured principles and practices to the development, implementation, and maintenance of software applications and systems that deliver functional solutions to meet specific user needs and business requirements \cite{bosch2015speed}.
Traditional SE processes rely heavily on human effort for critical tasks, such as requirements gathering, system design, coding, testing, and quality assurance \cite{liu2024large}. This human-centric approach has historically been a defining feature of conventional software development practices.

However, the recent integration of Artificial Intelligence (AI) into SE has led to significant advancements and opened up new possibilities for synthesizing software processes and practices with data/intelligence-driven tools and techniques to engineer software systems \cite{treude2023navigating}. In recent years, a growing number of studies have explored the role and impact of Large Language Model (LLM)-based agents in automating SE processes \cite{jin2024llms}. Modern SE processes follow an agent-centric approach, where agents play an important role in the software development process \cite{hassan2025agentic}. Despite the steady growth in published research on LLM-based agents as AI assistants in software development, empirical evidence on their effectiveness in practical development settings remains limited. In particular, most existing studies rely on benchmark-driven or automated evaluations, with little emphasis on practitioner-centric assessments that capture developers’ perspectives on the capabilities and limitations of these agents in real-world software development tasks.

\textbf{Motivation and Contribution}: Our study is part of the MAISA and BF/Amalia-SW projects, both funded by Business Finland. In collaboration with Finnish companies, this study aims to advance the SE field by utilizing LLM-based agents and evaluating their capabilities in real-world scenarios. This research addresses an industry need to understand the practical applicability of agents for software development, including their strengths, challenges, and areas for improvement. To address this need, we conducted an empirical study consisting of two main steps, as shown in Figure \ref{fig:multi agent system}: (i) the development of a multi-agent system, CodePori, designed to autonomously generate code for software systems from high-level descriptions, and (ii) the evaluation of the agents' performance, strengths, and limitations through a survey involving 118 participants.

\textbf{Contribution}: Our key contributions can be summarized as follows:

\begin{itemize}
    \item We have designed and developed a multi-agent system, CodePori, to support autonomous, project-level code generation from high-level descriptions, producing structured, multi-file software projects.
    \item We provide empirical insights into multi-agent systems for code generation based on feedback from 118 participants.
    \item We present a comprehensive analysis of the multi-agent system's performance, including its strengths, challenges, and areas for improvement, providing researchers and practitioners with valuable evidence to guide future research on agent-based systems in SE.

    \item To support replication, validation, and broader exploration of the findings of our study, we have publicly released the survey dataset \cite{CodePori_Dataset} and the proposed multi-agent system via GitHub \cite{CodePori_v02} to aid further research and implementation.
\end{itemize}


\textbf{Structure:} The rest of the paper is organized as follows. The related work is presented in Section \ref{sec:RelatedWork}. The study methodology is described in Section \ref{sec:Method}. The results are presented in Section \ref{sec:Results}, followed by a discussion in Section \ref{sec:Discussion}. Threats to validity are addressed in Section \ref{Thread to validity}. Finally, the paper concludes in Section \ref{sec:Conclusions}.

\section{Related Work}
\label{sec:RelatedWork}



Over recent years, LLMs have shown significant growth and potential in various SE applications, including code generation \cite{feng2023investigating}. Several code-oriented LLMs have been introduced to facilitate both code comprehension and generation tasks. For instance, prominent models include AlphaCode (Li \textit{et al}. \cite{li2022competition}), Codex (Chen \textit{et al}. \cite{chen2021evaluating}), CodeGeeX (Zheng \textit{et al}. \cite{zheng2023codegeex}), and InCoder (Fried \textit{et al}. \cite{fried2022incoder}). These models provide a foundation for advances in LLM-based software development by introducing key capabilities, such as code completion and code generation from natural-language descriptions. However, in 2022, the introduction of the ReAct framework by Yao \textit{et al}. \cite{yao2022react} marked an important shift toward LLM-based agent systems by enabling LLMs to interleave reasoning with task-specific actions. This capability provided a foundation for agents that can reason, act, and interact with external environments to accomplish complex tasks.


Based on this development, in 2023, Qian \textit{et al}. \cite{qian2023communicative} introduced the ChatDev system, which utilizes LLM-based agents throughout the entire software development life cycle. This was the first system to use LLM-based agents for software development.
Hong \textit{et al}. \cite{hong2023metagpt} introduced MetaGPT, an innovative meta-programming framework that incorporates efficient human workflows into LLM-based multi-agent collaborations. This framework achieved 89\% accuracy on HumanEval and 87.7\% on the MBPP benchmark \cite{hong2023metagpt}. MetaGPT uses an assembly line paradigm to assign diverse roles to various agents, efficiently breaking down complex tasks into sub-tasks.
Following MetaGPT, Huang \textit{et al}. \cite{huang2023agentcoder} introduced AgentCoder, a multi-agent framework designed to improve the quality and efficiency of code generation by simulating a collaborative team of developers. AgentCoder achieved 96.3\% on HumanEval and 91.8\% on the MBPP benchmark, demonstrating higher accuracy compared to both MetaGPT and ChatDev.
In 2024, Islam \textit{et al}. \cite{islam2024mapcoder} proposed MAPCoder, an agent-based system focused on improving performance by breaking down large problems into smaller, manageable sub-tasks and using a structured communication protocol between agents to ensure a coherent final solution. This framework achieved 93.9\% on HumanEval and 83.1\% on the MBPP benchmark, showing an improvement over MetaGPT, especially on the HumanEval benchmark. However, AgentCoder’s performance of 96.3\% and 91.8\% on HumanEval and MBPP, respectively, still shows higher accuracy compared to MetaGPT, ChatDev, and MAPCoder.
Most recently, SWE-Agent \cite{yang2024swe} is an agent-based framework designed to solve software issues from GitHub repositories. It uses a specialized agent that can autonomously execute code, navigate repositories, and apply fixes, demonstrating high accuracy compared to the models mentioned above.

However, as we have observed, while current agent-based systems show high accuracy on benchmarks, their performance on real-world problems is often low \cite{paul2024benchmarks}. As noted by Yadav \textit{et al.} \cite{yadav2024boldly}, these benchmarks contain a limited set of projects and do not represent the full range of real-world programming challenges. Similarly, Dai \textit{et al.} \cite{dai2024mhpp} pointed out that the programming tasks in the HumanEval dataset are designed for small projects. In particular, benchmark accuracy primarily reflects performance on isolated tasks and does not capture system-level challenges, such as agent capabilities, inter-agent coordination, and reliability across multi-step development workflows. As a result, strong benchmark performance does not necessarily translate into effective system behavior in realistic software development settings \cite{mohammadi2025evaluation}. This represents a notable gap in evaluating systems intended to generate code for large-scale systems. To address this gap, our study extends existing work by involving participants to examine the effectiveness and practical limitations of multi-agent systems that are not captured by traditional benchmarks. Building on existing multi-agent software development approaches, CodePori focuses on project-level, multi-file software generation and evaluates its practical applicability using participant-provided software projects.

\section{Research Method}\label{sec:Method}
This section presents the methodology used to conduct this study. Section \ref{RQs} states the research questions, Section \ref{System Design} describes the multi-agent workflow and roles, and Section \ref{Evaluation Framework} outlines the evaluation framework. The overall research method is illustrated in Figure \ref{fig:multi agent system}.

\subsection{Research Questions}
\label{RQs}
Based on the study goal, we formulated the following two Research Questions (RQs). 

\textbf{RQ1.} \textit{Can the multi-agent system autonomously generate code, and is it capable of handling the generation of large-scale systems?}

\textbf{RQ2.} \textit{What are the perceived effectiveness and practical limitations of the multi-agent system for autonomous code generation, as experienced by participants?}

\textit{\textbf{Objective.}} 
The main goal of \textbf{RQ1} is to determine if the multi-agent system can perform autonomous code generation for large-scale systems. The objective of \textbf{RQ2} is to examine the effectiveness and practical limitations of the CodePori system in autonomous code generation, achieved by analyzing participants’ feedback on the system’s performance in real-world scenarios.

\subsection{Multi-Agent System Design}
\label{System Design}

\begin{figure}[ht]
    \centering
    \includegraphics[width=\linewidth]{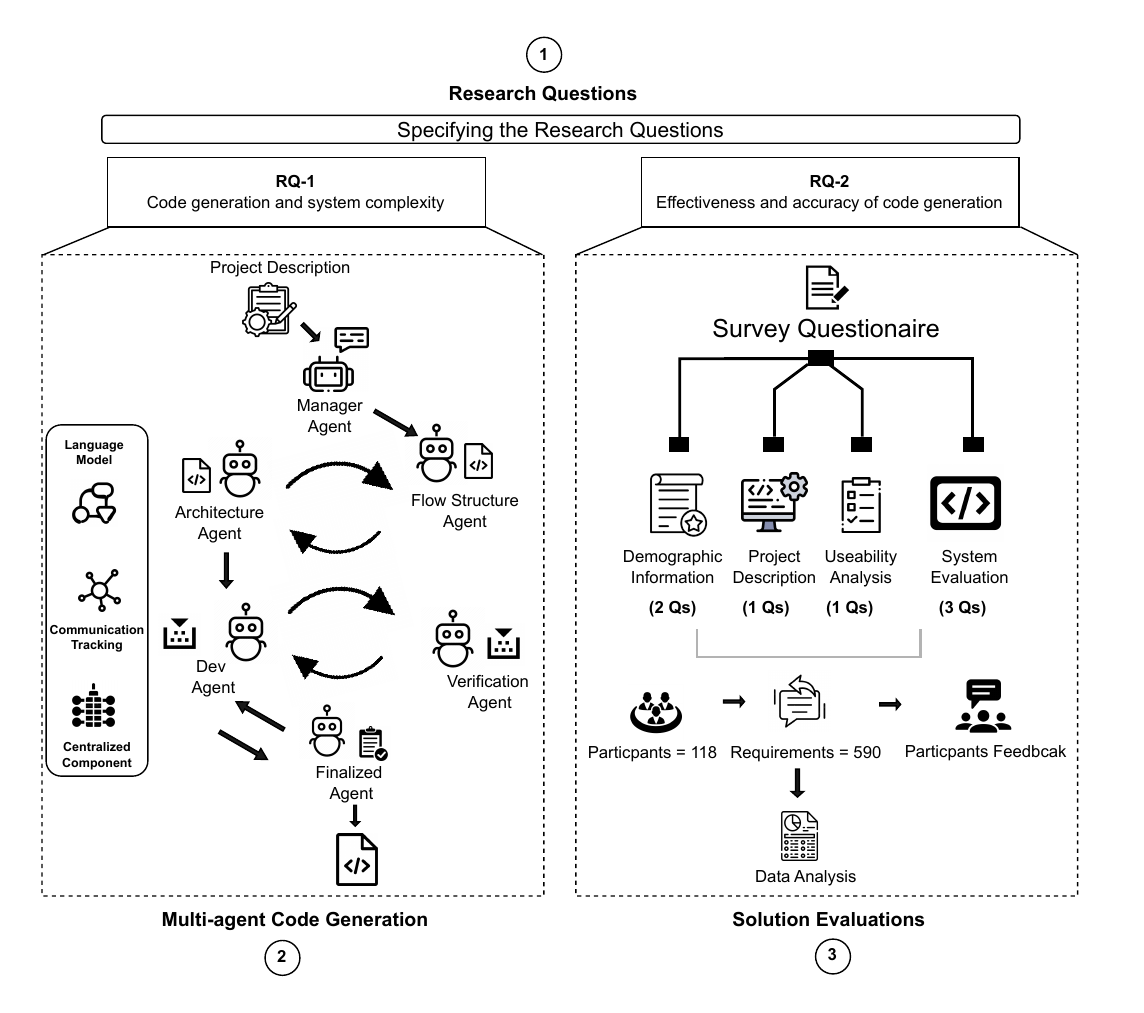}
    \caption{Multi-agent system and survey workflow.}
    \label{fig:multi agent system}
\end{figure}

As shown in Figure \ref{fig:multi agent system}, we designed six LLM-based agents that collaborate to autonomously perform coding tasks based on the project description. CodePori is designed as a research prototype for large-scale, project-level software generation, where multiple specialized agents collaboratively transform high-level requirements into structured, multi-file software projects. According to Levy and Feitelson \cite{levy2021understanding}, a large-scale system includes advanced architectural design, a multi-file modular structure, and intricate inter-component communication. In this study, we utilized instruction-based prompts following established prompting guidelines \cite{ouyang2022training}, which involve identifying the role, explaining the task, and defining the output format for each agent. Each agent specializes in specific tasks and interacts with others through APIs, facilitating continuous communication and refinement of the generated code. 
In the following, we provide a detailed explanation of each agent.

The \textbf{Manager agent} plays an important role in the autonomous code generation process by serving as a project manager responsible for analyzing the high-level project description. The \textit{Manager agent} is tasked with breaking down the project into structured components. This is achieved by processing the provided project description and identifying the necessary modules for development. The \textit{Manager agent} ensures a structured decomposition by requiring each module to be described in a standardized format, including attributes, such as module name, detailed description, objectives, expected inputs, expected outputs, dependencies, additional notes, and best practices. 


The \textbf{Architecture agent} plays an important role in the autonomous code generation pipeline by transforming modular specifications into a comprehensive architectural overview. The \textit{Architecture agent's} primary task is to provide a detailed, standardized architectural overview that captures how different components interact, manage data flow, and connect with the underlying databases. 


The \textbf{Flow structure agent} transforms high-level architectural planning into a structured file and folder hierarchy aligned with the selected programming language and tech stack. It interprets the project’s modular descriptions and organizes them into logically grouped directories, ensuring that configuration files, dependency managers, and code components are properly structured. The \textit{Flow structure agent} outputs a plain-text layout that includes key elements, such as main.py, config.py, and subdirectories.
By converting abstract architecture into a concrete file structure, the \textit{Flow structure agent} enables the other agents to generate and place code systematically. 

The \textbf{Dev agent} is responsible for generating initial code for each module or file identified by the \textit{Flow structure agent}. The \textit{Dev agent} ensures that each module is developed without placeholders or TODOs, aligns with previously completed modules, and follows Google-style docstrings for maintainability. Additionally, each module contains between 150 and 200 lines of code. To ensure consistency and quality, the \textit{Dev agent} ensures that the code follows predefined coding guidelines and industry best practices, including optimized algorithm selection, structured exception handling, and efficient memory management. 

The \textbf{Verification agent} is responsible for conducting a detailed code analysis to ensure that the current module code aligns with the project description and that all modules interact correctly. This agent does not modify the code directly, but instead sends it back to the \textit{Dev agent}, accompanied by clear instructions on the necessary improvements to enhance the system's functionality. The \textit{Verification agent} plays a strict role in code review, ensuring the highest code quality. 


The \textbf{Finalization agent} specializes in the refinement and enhancement of the current module code to ensure that it is production-ready. The process starts with receiving the project description, file and folder structure, generated code, and a review report detailing the necessary improvements. The \textit{Finalization agent} first analyzes the existing implementation, understands the required modifications, and then collaborates iteratively with the other agents to enhance the code quality. The \textit{Finalization agent} ensures that the final code maintains high integrity, follows industry best practices, and aligns with the overall architectural requirements of the project. 
\subsubsection{Core Components}
\label{Core Components}
The proposed system consists of the following core components:
 \begin{itemize}
     \item \textbf{Language Model:} In this project, we used the Claude-3.5 Sonnet model for code generation. We chose this model because it outperforms previous state-of-the-art models on the HumanEval benchmark, achieving 92\% accuracy. 
     We used an API key to connect to the system, with API requests made during each iteration of the six-agent discussions for collaborative code generation.
     
     \item \textbf{Agent Communication Tracking:} We developed a structured environment, which includes a specialized logger to track all agents’ communication activities. This initial configuration ensures that all operations, from initial project design to final code generation, are performed within a controlled and traceable environment.
     
     \item \textbf{Centralized Component:} To facilitate coordination and data persistence across agents, a state manager was created. This centralized component maintains critical state information, such as the project description, architectural overview, folder structure, all generated code files, and verification reviews. The state manager ensures that each agent has access to the necessary context and accumulated results from previous steps, thereby enabling smooth collaboration between agents and minimizing redundant processing.
     
    \item \textbf{Agent Orchestration Pipeline:} The pipeline coordinates the end-to-end process between six agents, ensuring that the software project is developed in a modular and verifiable manner, utilizing the strengths of LLM-based agents at each critical stage of the pipeline.
     

 \end{itemize}

\subsection{Survey Design}
\label{Evaluation Framework}
Concerning the research goals, we selected a survey as the primary data collection method to understand participants’ perceptions of the proposed multi-agent system for code generation. We followed the survey guidelines provided by Linaker \textit{et al}. \cite{linaaker2015guidelines}. In the evaluation phase, participants were asked to assess the system’s effectiveness, practical limitations, and real-world applicability.

\subsubsection{Survey Questionnaire}
We created the survey in English using Microsoft Forms, consisting of both open- and closed-ended questions. As depicted in Table \ref{tab:survey-questions}, the first two Survey Questions (SQs) were about the participants’ background, covering their education and experience in software development and LLMs. In the second part, we designed five questions (SQ3--SQ7) aimed at understanding their perspectives on the use of multi-agent systems for large-scale code generation.

\begin{table}[h]
\centering
\caption{Survey questions}
\label{tab:survey-questions}
\small
\begin{tabular}{|c|p{0.78\columnwidth}|}
\hline
\textbf{ID} & \textbf{Questions} \\ \hline

SQ1 & What is the highest level of education you've completed? \\ \hline

SQ2 & How much experience do you have in software development and LLMs? \\ \hline

SQ3 & Write the five projects you provided for code generation. \\ \hline

SQ4 & How long does the system take to generate code for a single project description? Please write the time taken for each project. \\ \hline

SQ5 & Is it required to modify the code generated by the AI agent system? If yes, what modifications did you make to execute the code? Please write the modifications you made for each project. \\ \hline

SQ6 & How do you rate the proposed AI agent system for code generation? \\ \hline

SQ7 & Are you satisfied with the performance of the proposed system? Please elaborate on its strengths and areas that need improvement. \\ \hline

\end{tabular}
\end{table}

\subsubsection{Participant Recruitment}
To evaluate the effectiveness, practical limitations, and real-world applicability of the multi-agent system, we engaged 118 students enrolled in the ``Fine-Tuning LLMs'' course held at Tampere University in 2025. Aside from practical assignments where the students fine-tuned open-source LLMs, the course included teaching on other topics related to LLMs and SE, such as LLM-based AI agents.

In the final lecture of this course, we introduced our system to these 118 students. We first provided a detailed introductory explanation of the proposed system, and then demonstrated its functionality in practice for the students. This was meant to ensure that the students had an understanding of the system's functionality before participating in the evaluation. We then shared the survey link together with the GitHub repository of the proposed system, along with detailed instructions on how to use the system. We acknowledge that the use of student subjects for the evaluation poses some threats to the validity of our study. We discuss these in detail in Section \ref{Thread to validity}. However, we do note that, having taken this course before participating in the evaluation, these students had verifiable prior experience with LLMs through the course itself.




\subsubsection{Analyzing Survey Data}

Descriptive statistics \cite{wohlin2006empirical} were used to perform a quantitative analysis of the closed-ended questions. Moreover, we applied the open-coding method \cite{blair2015reflexive} to analyze the open-ended response data. The first author conducted the open coding and subsequently used selective coding to identify the core categories by aggregating related concepts. To reduce personal bias during data analysis, the second author participated in validating the generated codes.

\section{Results}\label{sec:Results}

This section presents the study results. Section \ref{codePori} describes the implementation and demonstration of CodePori. Section \ref{Participants evaluation result} reports on the survey findings from the 118 participants who evaluated the system.

\subsection{CodePori: LLM-Based Multi-Agent System (RQ1)}
\label{codePori}



 \begin{figure*}[ht]
    \centering
    \includegraphics[width=0.9\linewidth]{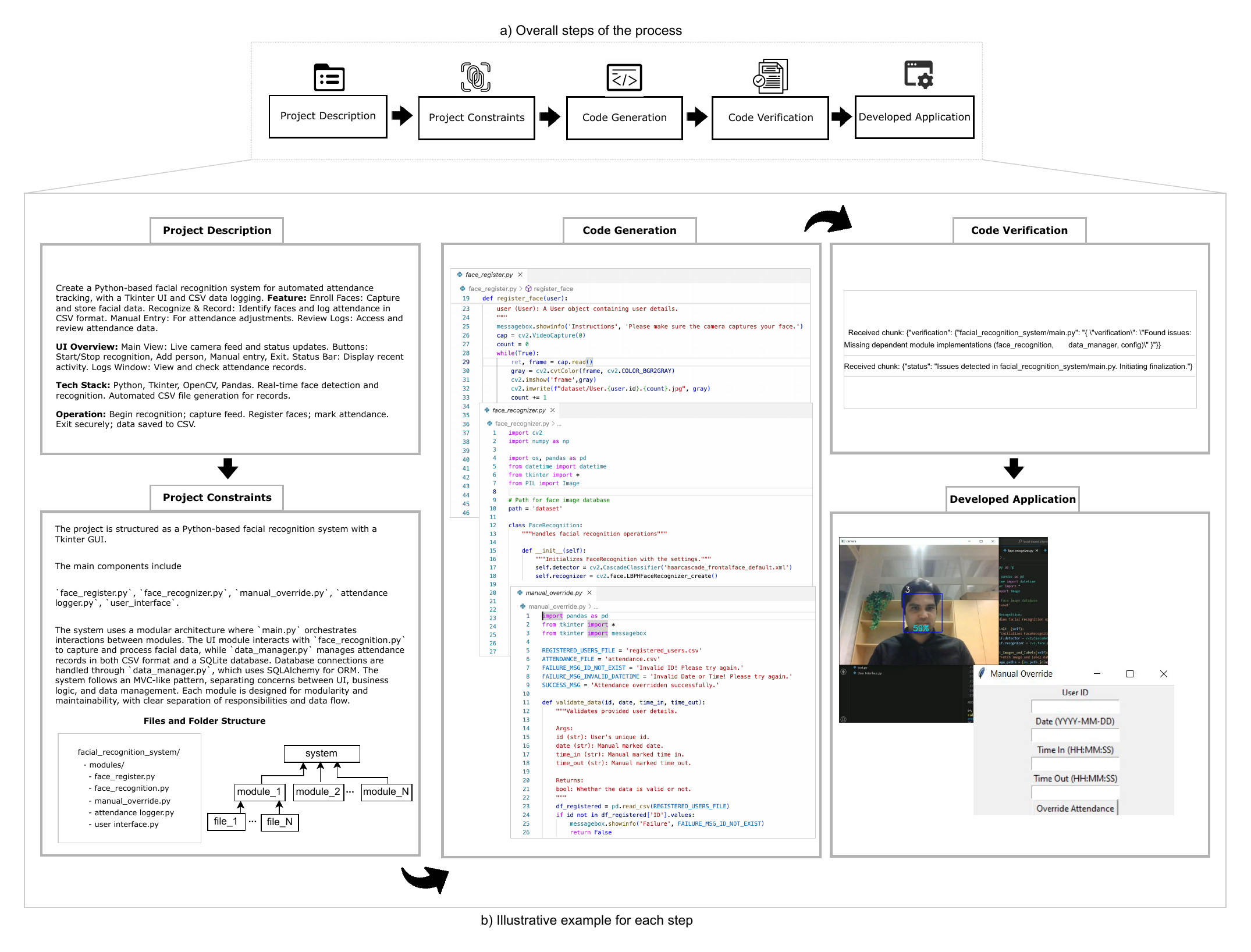}
    \caption{Detailed representation of the software system developed by CodePori.}
    \label{fig:placeholder}
\end{figure*}


The results indicate that LLM-based agent systems show potential for autonomous, project-level code generation involving structured, multi-file software systems; however, manual modifications were often required to achieve successful execution, particularly for more complex projects. Figure \ref{fig:placeholder} illustrates how six agents collaborate to transform high-level requirements into a complete software application. The process begins with a detailed project description provided as input. The first agent interprets the requirements and identifies five modules, each accompanied by a description of its functionality. Two subsequent agents then define the system architecture and establish a structured file and folder hierarchy. Based on this architecture, the system generates the initial code for the identified modules. The generated code is then reviewed by a verification agent, which provides feedback for improving its consistency and correctness. The final output is a complete software application.
Despite this largely autonomous process, some manual modifications were required to ensure successful execution of the generated application. As shown in Figure \ref{fig:placeholder}, these modifications included creating or configuring the required project files, updating file paths, and adding necessary external resources or dependencies.

Overall, our initial findings demonstrate the use of multi-agent systems to support end-to-end software development with minimal human involvement. However, human--agent interaction remains important for addressing execution-specific issues and ensuring the reliability of the final application.

\subsection{Survey Results (RQ2)}
\label{Participants evaluation result}

We collected feedback from 118 participants to evaluate the proposed multi-agent system for autonomous code generation. In total, 590 project descriptions submitted by the participants were analyzed and classified to assess how the system handled different types of project inputs. 
The analysis also included measuring the average completion times across project categories to assess system efficiency. Furthermore, we evaluated the overall performance of the system by identifying its key strengths, existing challenges, and areas for improvement based on participant feedback, focusing on the effectiveness of multi-agent collaboration in code generation tasks.


\subsubsection{Demography of Participants (SQ1-SQ2)}

Below, we outline the demographic information to provide a clearer overview of the participants’ characteristics.

\begin{itemize}
   \item \textbf{Participants' experience}: 82 out of 118 participants (69.49\%) had around 2--5 years of experience in AI-based software development, 31 participants (26.28\%) had approximately 2 years of experience, and 5 participants (4.23\%) had around 10 years of experience in this domain.
\end{itemize}

\begin{itemize}
    \item \textbf{Qualifications}: The majority of participants, i.e., 56 out of 118 (47.46\%), held a bachelor's degree, followed by 33 out of 118 (27.97\%) with a master's degree, 26 out of 118 (22.03\%) with a high school diploma, and only 3 out of 118 (2.54\%) with a doctoral degree.
\end{itemize}

\subsubsection{Project Description (SQ3)}
In this survey, each participant was asked to provide five different project descriptions as input. In total, 590 descriptions were collected to test the system's capabilities. Participants were free to choose any application domain; therefore, the resulting categories reflect participant-selected domains rather than a predefined set. We analyzed and classified the descriptions to understand how the multi-agent system responds to varying levels of detail and project scope. As shown in Table \ref{tab:project_classification}, the descriptions were grouped into 11 categories: Games, AI/ML, Finance, 2D/3D Visualization, Chatbots, Benchmarking, Mathematical, Image Processing, E-commerce, Medical, and Others. This classification also allowed us to examine the applicability of the multi-agent system across a diverse range of software application domains. We further examined how the system performs when given vague project descriptions versus highly detailed and specific requirements.
\begin{table}[h!]
\centering
\caption{Classification of project descriptions}
\label{tab:project_classification}
\begin{tabular}{|l|c|c|}
\hline
\textbf{Category} & \textbf{\# of Projects} & \textbf{Percentage (\%)} \\ \hline
Games & 149 & 25.25\% \\ \hline
AI/ML & 48 & 8.14\% \\ \hline
Mathematical & 50 & 8.47\% \\ \hline
Finance & 24 & 4.07\% \\ \hline
Image Processing & 7 & 1.19\% \\ \hline
E-commerce & 7 & 1.19\% \\ \hline
Chatbots & 26 & 4.41\% \\ \hline
2D/3D Visualization & 6 & 1.02\% \\ \hline
Medical & 6 & 1.02\% \\ \hline
Benchmarking & 1 & 0.17\% \\ \hline
Others & 266 & 45.08\% \\ \hline
\textbf{Total Projects} & \textbf{590} & \textbf{100\%} \\ \hline
\end{tabular}
\end{table}

\subsubsection{Required Modification (SQ5)}
Participants reported that the generated code often required manual modifications before successful execution, particularly for projects involving multiple files and more complex project structures. The required changes commonly included correcting import paths, resolving inconsistencies between modules, adding missing files or functions, fixing dependency and database configuration issues, and addressing incomplete code generation. In contrast, simpler projects were more likely to execute with little or no modification.

\subsubsection{Overall Performance (SQ6--SQ7)}
We define overall performance in terms of the system's speed (i.e., the time required to generate and complete the code) and code-generation accuracy. In this study, 30.51\% of participants rated the system as poor, 20.33\% rated it as good, and the remaining 49.16\% rated it as average or fair (26.28\% and 22.88\%, respectively).




\paragraph{Strengths (SQ7)}
Participants highlighted six key strengths of the proposed system for autonomous code generation. These include fast code generation, a clear understanding of user requirements, code comments and documentation, efficiency and ease of use, a multi-layer architecture, and suitability for prototype-level projects. A detailed discussion of these identified strengths is provided below.

\textbf{Fast code generation}: To evaluate the speed of the proposed system, we performed a descriptive analysis on SQ4, examining the average completion time across 590 projects. The results indicate that the average completion time with the system was approximately 9 minutes 8 seconds, based on a total of 5,391 minutes across 590 data points. This suggests that, on average, the system enables relatively quick project turnovers. Additionally, 10 out of 118 participants (8.47\%) highlighted that the proposed system reduces development time by handling various aspects of coding. For instance, a participant highlighted that “\textit{this AI agent code generation tool has shown strong capabilities in automating the development process by generating structured and functional code. It can effectively reduce development time by handling various aspects of coding, including routing, database integration and user interface components, even for complex project descriptions.}” (\textbf{Participant 65}).



\textbf{Better requirement interpretation}: In this study, 3 out of 118 participants (2.54\%) highlighted that the proposed system clearly understands the provided requirements and effectively translates them into well-structured and functional code outputs. For example, one participant highlighted that “\textit{i'm very satisfied as personally I wouldn't have achieved structure and professionalism in code generation without these tools. The strength of the system is that it very clearly understands what is asked and what the user wants to be done.}” (\textbf{Participant 46}).


\textbf{Code comments and documentation} are important aspects of the software development process, as they enhance code readability, maintainability, and collaboration among developers. In this study, 2 out of 118 participants (1.69\%) highlighted that the proposed system effectively generates well-structured and concise code comments, as well as supporting documentation. For instance, a participant mentioned that “\textit{yes, I am satisfied with the performance of the proposed system. Its first strength is that it gave well-written and concise comments, both the code block comment and the function comment, also the program documentation.}” (\textbf{Participant 101}).


\textbf{Efficiency and ease of use}: In this study, 12 out of 118 participants (10.17\%) emphasized that the proposed system was efficient in automating setup tasks, managing dependencies, and organizing the project structure. Participants' feedback highlights that the system accelerates the development process and improves ease of use by simplifying the initial setup and ensuring a clean, well-structured workflow. For instance, a participant stated that “\textit{yeah, it actually works way better than I expected! The way it automatically sets up the whole directory and installs all the required dependencies is super convenient. It saves a lot of time and effort, making it easy to get started without worrying about setup hassles.  I also love how it keeps everything well-organized, making the project structure clean and easy to navigate.}” (\textbf{Participant 42}).


\textbf{Multi-layer architecture} refers to the structured design of software systems into distinct layers, where each file is responsible for specific functionalities. In this study, 3 out of 118 participants (2.54\%) highlighted that the proposed system generates projects with a clear multi-layer architecture, making development easier, less error-prone, and more accessible even to non-coders. As an example, one participant noted that “\textit{one strength is that the system has a multi-layer architecture, which allows for different perspectives on the project. Both the front and back end of the project can be completed with little to no bugs, which saves much of the human effort needed for development. This allows for even non-coders or people with little technical background to be able to develop applications easily and quickly.}” (\textbf{Participant 52}).




\textbf{Good for prototype-level projects}: In this study, 6 out of 118 participants (5.08\%) mentioned that the proposed system works efficiently for prototype-level or small-scale projects. However, they also noted that the system fails to handle more complex projects, where errors become more frequent and persistent. For instance, a participant highlighted that “\textit{the system handles simpler projects efficiently, but it gave persistent failure in generating complex project structures. It mostly had finalizer bot error and stopped.}” (\textbf{Participant 24}).


\paragraph{Challenges (SQ7)}
\label{Challenges}
In this study, 36 out of 118 (30.51\%) participants reported challenges with the agent-based system for autonomous code generation. Participants highlighted several challenges, such as hallucination, short-term memory, code smell, poor code quality. Each challenge is explained in detail below.

\textbf{Hallucination}:
According to Maleki \textit{et al}. \cite{maleki2024ai}, hallucination refers to the model's generation of content that is factually inaccurate, illogical, or entirely fabricated, despite appearing reasonable and coherent. In this study, 14 out of 118 (11.86\%) of the participants pointed out that agents hallucinated by generating unrequested code components, introducing unnecessary functionalities, or misinterpreting project requirements by including elements not aligned with the specified description (e.g., adding unwanted dependencies or architectural choices). For instance, a participant stated that “\textit{generated code tend to have more functions than requested, sometimes making code unusable without additions. Like database situation where model hallucinates need of database although application or descriptioned functionality doesn't need one.}” (\textbf{Participant 46}).




\textbf{Short-term memory}:
LLM-based agents' short-term memory has emerged as a significant limitation; for instance, as tasks become more complex, previous information tends to be forgotten. In this study, 7 out of 118  (5.93\%) participants highlighted that agents often forget the previously generated class or function in another module when working with large-scale systems, leading to errors, such as calling non-existent methods or failing to maintain inter-file references following code refactoring. As an example, one participant noted that “\textit{there are couple of issues that makes me wonder whether the AI actually has full understanding of the whole system. For example calling non-existent methods. I believe this could be caused by a phenomena where file x is using file y's methods but then file y is refactored due to failed tests and the AI forgets that the methods were referred in file x.}” (\textbf{Participant 25}).


\textbf{Code smell} is not a bug, but it is a structural or stylistic characteristic in the source code that indicates a potential deeper problem or suboptimal design, making the code harder to understand, modify, or maintain \cite{guggulothu2020code}. In this study, one participant raised the code smell issue with the proposed system's generated code. The participant highlighted specific instances of poor code quality, such as the hardcoding of constants in multiple locations, extensive duplication of code segments, and the presence of dead or unused code. The participant highlighted that “\textit{code quality was also often very poor, with the same constants hardcoded in several places and lots of duplicated code and dead code.}” (\textbf{Participant 118}).


\textbf{Code inconsistency}: In this study, 4 out of 118 (3.39\%) participants identified code inconsistency issues in the proposed system's generated code. Specifically, they highlighted the absence of uniform patterns in function calls and variable definitions across generated files. For instance, a participant highlighted that “\textit{the system does provide codes that offers an overall framework and all the basic functionalies in related to the project description, but the generated codes never work out of the box, there are a lot of inconsistences for function calls and variable definitions across generatedpython files, regardless of the complexity of the projects.}” (\textbf{Participant 57}).


\textbf{Poor code quality}: In this study, 20 out of 118 (16.95\%) participants mentioned that the generated code quality of the proposed system was bad. The generated code frequently suffered from functional defects and dependency conflicts, which made it unusable without significant debugging and modifications. For instance, a participant stated that “\textit{no, messy code and hard to debug. Considering the quality vs time spent coding, I still think that coding them by yourself is better.}” (\textbf{Participant 61}).



\textbf{Dependency management issue}:
According to Pashchenko \textit{et al}. \cite{pashchenko2020qualitative}, dependency management issues arise when software systems improperly handle, track, or maintain their reliance on external libraries, frameworks, modules, or services. In this study, 2 out of 118 (1.69\%) participants raised dependency management issues with the code generated by the proposed system. For instance, a participant highlighted that 
“\textit{it gave pretty good skeleton for a program, but the programs dont work straight out of the package, there is too much errors and problems there. Dependencies had conflicting versions that had weird errors etc. Then module errors so much.}” (\textbf{Participant 59}).


\textbf{Syntax error} occurs when the grammatical rules or structural conventions of a programming language are violated, which prevents the code from being correctly understood or processed by a compiler or interpreter. In this study, 3 out of 118 (2.54\%) participants highlighted that the code generated by the proposed system often suffered from syntax errors. As an example, one participant noted that “\textit{most of the cases It can not generate the full code. When it generates the full code, those section needed to update as well, syntax error and import required libraries or modules are some common problem I faced with the generated code.}” (\textbf{Participant 82}).


\textbf{Orchestration failure}:
Agent orchestration is the automated coordination of multiple AI agents to ensure that specialized agents can work together in a cohesive workflow to achieve a complex goal \cite{xiong2025self}. In this study, 2 out of 118 (1.69\%) participants highlighted agent orchestration issues between different files. As illustrated by one participant, “\textit{at least from what I noticed, it doesn't seem to know how to orchestrate between different files. It has a tendency of creating a requirements-file but not actually adding the requirements (also other excess files).}” (\textbf{Participant 49}).


\textbf{Code complexity}:
According to Geraci \cite{geraci1991ieee}, code complexity is defined as ``the degree to which a system or component has a design or implementation that is difficult to understand and verify, which directly affects the maintainability and defect proneness of code.'' In this study, 6 out of 118 (5.08\%)
participants highlighted the code complexity issue with agent-based generated code. One participant highlighted this point by stating that “\textit{however, due to the complexity of the project structure and processing logic, there are inevitably some bugs, which require careful analysis and one-by-one processing by humans. The larger the project, the more complex the associations, and the more difficult it will be for the system to perform.}” (\textbf{Participant 103}).


\textbf{Configuration issue}:
In this study, 1 out of 118 participants (0.85\%) identified system integration and configuration as key challenges, specifically highlighting difficulties in getting generated files to work cohesively and ensuring compatibility between different components.
For instance, a participant highlighted that “\textit{importing seems to be very problematic for the system to get right and sometimes it uses non-compatible methods in different files. As I mentioned before, the biggest limitation seems to be to get the generated files to work with each other.}” (\textbf{Participant 45}).



\paragraph{Areas of Improvement (SQ7)}
\label{Area of improvement}

Based on real-world testing of the proposed system, participants suggested several areas for improvement. Below we discuss each area of improvement in detail.

A \textbf{progress tracking} system provides users with real-time updates and visual indicators regarding the status and progress of an ongoing process or task. This enables users to monitor efficiency and identify any potential delays or issues within the system's operation.
In this study, 8 out of 118 (6.78\%) participants suggested the addition of a progress tracking system, emphasizing its importance for users to monitor the agent's progress. Such a feature would enhance user experience by providing critical feedback on long-running or complex operations. For instance, one participant highlighted that “\textit{for the python app in general I would like to see some kind of progress bar to better gauge if the app has gotten stuck during execution or not.}” (\textbf{Participant 50}).

\textbf{Improve code debugging}:
In this study, 14 out of 118 (11.86\%) participants argued that the system should provide comprehensive instructions for all types of errors across all relevant files. Furthermore, there should be a separate debugging option that allows users to provide a file containing an error for agents to analyze and resolve. As an example, one participant noted that “\textit{The idea is very good and for the most part the code runs well. There should be a place to debug, or give instructions for debugging. That would be a nice add.}” (\textbf{Participant 18}).

\textbf{Human intervention with agents} is crucial for improving performance and overall effectiveness. In this study, one participant argued that the system lacked human intervention when agents were developing code or crashed at some point, meaning that humans were unable to step in to diagnose, guide, or rectify issues, leading to frustration and the inability to resolve problems effectively. For instance, a participant mentioned that “\textit{One problem is that the whole system crashes when the agents face errors, and as a user you cannot intervene at all. This was many times the case and since there is no detailed error-messages, I have no idea how to fix the bug or what went wrong.}” (\textbf{Participant 97}).



\textbf{Framework selection}:
In this study, two participants suggested that JavaScript framework detection should be improved to generate vanilla JavaScript by default and only use Node/React when explicitly requesting framework-based projects. For instance, a participant stated that “\textit{for javascript it generates code using node and react if I mention that the code is required for js. I have to explicitly mention that I just need it for vanilla javascript then it generates the required code. It should generate code using node or react when I mention that I want a project using javascript's framework.}” (\textbf{Participant 99}).


\section{Discussion}\label{sec:Discussion}


\subsection{LLM-Based Agent Code Generation (RQ1)}
The implementation of LLM-based agents in software development demonstrated their potential for autonomous, project-level code generation involving structured, multi-file software systems. However, manual modifications were sometimes required to achieve successful execution, indicating that human involvement remains necessary in the development process. 
Several researchers have explored the use of multi-agent systems for code generation. For instance, AgentCoder \cite{huang2023agentcoder}, SWE-Agent \cite{yang2024swe}, MetaGPT \cite{hong2023metagpt}, and similar frameworks utilize multi-agent collaboration to support software development. While some existing approaches support collaborative or end-to-end software development, many evaluations still rely on isolated coding tasks or benchmark-level code generation. Building on this line of work, our proposed LLM-based AI agent system focuses on project-level, multi-file code generation.
\textbf{Implications}: For practitioners, the use of LLM-based agents can reduce development time and assist in software development tasks, allowing them to focus on the more complex and creative aspects of their projects. This can lead to increased productivity and efficiency within development teams. Additionally, the ability to generate code autonomously can result in cost savings, as projects are completed more efficiently and with fewer resources \cite{hong2023metagpt}. From an academic and research perspective, this advancement enhances the understanding of AI-driven software development by improving the capabilities and efficiency of autonomous code generation. Furthermore, LLM-based agents create opportunities for future research, such as examining their impact on the SE workforce, studying the implications of AI-generated code, and developing frameworks to ensure transparency, accountability, and reliability in AI-assisted programming systems. Finally, enhancing the system’s learning capabilities based on user input and customizing its outputs according to specific user preferences or project requirements could further improve its usefulness.

\subsection{Performance Evaluation of LLM-Based Agents (RQ2)}
The study highlights the strong potential of LLM-based agents while also identifying several challenges, as discussed in Section \ref{Challenges}. These challenges represent key empirical findings of this work, as they provide valuable insights into the current challenges faced by multi-agent systems. Below, we discuss these challenges and their implications for practitioners and researchers.

\textbf{Implications for practitioners:}
The findings provide practical guidance for practitioners seeking to adopt LLM-based multi-agent systems in real-world software development and industrial settings. For instance, human oversight, continuous monitoring, and structured review processes are important for ensuring reliability and preventing incorrect or misleading code from being incorporated into production systems. Relying only on agents’ outputs without verification can be harmful to software projects. In addition, the identified code smells and code inconsistencies highlight the importance of establishing clear coding standards and coordination mechanisms when multiple agents operate within a shared development environment. Without unified guidelines for naming conventions, documentation, and modular design, generated code can diverge in structure and style, reducing its maintainability. Therefore, agents should be treated as contributors operating within a well-defined organizational and architectural framework rather than as independent developers. Finally, monitoring dashboards and orchestration controllers can help practitioners detect coordination breakdowns at an early stage.

\textbf{Implications for researchers:} The identified challenges provide several opportunities for future research on LLM-based multi-agent systems. For instance, the existing challenges (see Section \ref{Challenges}) and proposed areas for improvement (see Section \ref{Area of improvement}) require further investigation to enhance the reliability and effectiveness of these systems. Empirical studies are also needed to examine how different coordination strategies influence code quality and team productivity in agent-based development. With respect to orchestration failure, future research can explore hybrid orchestration approaches that combine symbolic planning with LLM-based reasoning to support more reliable collaboration and improve agent communication for complex software engineering tasks. Finally, the limitations of existing benchmarks indicate a need for evaluation frameworks that go beyond simple, single-function code generation. Such benchmarks should include complex, real-world tasks requiring system-level understanding, multi-file management, and interaction with external APIs or databases, while also evaluating code-quality aspects, such as readability, maintainability, and security.

\section{Threats to Validity}
\label{Thread to validity}

We followed the guidelines proposed by Runeson and Höst \cite{runeson2009guidelines} to discuss the threats to validity.

\subsection{Internal Validity}


LLM-based agents introduced two potential threats to internal validity. First, LLM-based systems are non-deterministic, meaning that the same prompt may produce different outputs across multiple runs. To mitigate this threat, each participant was assigned five projects, allowing us to obtain a more stable and representative view of the system’s performance. Second, recruiting participants with extensive knowledge of LLMs, software development, and agent-based systems is inherently challenging. Therefore, our participant pool consisted of individuals who had completed the ``Fine-Tuning LLMs'' course and had varying backgrounds, levels of experience, and technical skills. While this selection ensured a baseline level of familiarity with LLMs and related technologies, the participants were primarily students, which may limit the extent to which their evaluations reflect those of experienced industry practitioners. However, we note that our questions (Table \ref{tab:survey-questions}) were not ones that would require very specific types of SE knowledge or expertise, and were instead quite general in nature. We thus refer to past literature that argues that the suitability of student subjects largely depends on the task at hand and the expertise required (e.g., \cite{host2000using}), and argue that for this evaluation, students would be reasonably analogous to industry professionals in terms of being able to provide a general evaluation of the system.

\subsection{External Validity}


In this study, we used the Claude-3.5 Sonnet model because it was the most advanced model available when the experiment was conducted. Since then, more capable LLMs have been released, which may exhibit different performance characteristics. In addition, more recently developed agentic frameworks offer advanced external tool capabilities, such as web search, retrieval, and automated tool integration; however, these capabilities were not available during the development and evaluation of our agent system. These advancements should therefore be considered when assessing the generalizability of our findings. Furthermore, although we recruited a large number of participants, all of them were affiliated with a single university. This may limit the generalizability of our findings and may not fully represent the perspectives of LLM practitioners in other contexts.

\subsection{Construct Validity}


A potential threat to construct validity arises from how participants evaluated the quality of the generated code. Participants’ interpretations of code quality varied: some viewed successful execution as a sufficient indicator of high performance, while others considered broader aspects, such as readability, structural quality, or technical debt. This variation between participants’ interpretations and the broader construct of software quality introduces ambiguity regarding how accurately the evaluation reflects the system’s true capabilities. In addition, the evaluation primarily relied on participant-reported assessments, while objective measures, such as build success, test pass rates, functional correctness, and execution without modification, were not systematically collected. Moreover, the varying levels of detail in SQ5 responses limited our ability to quantify how many projects required modification before execution.

\subsection{Conclusion Validity}

A potential threat to conclusion validity is the subjective interpretation of the survey responses during data analysis. To mitigate this threat, the first author conducted the initial analysis of the survey data, while the second author independently reviewed the analysis to ensure consistency with the intended constructs.

\section{Conclusion and Future Work}
\label{sec:Conclusions}
In this study, we investigated the real-world applicability of LLM-based agents for software development. Our findings provide useful insights for both researchers and practitioners by highlighting the current capabilities, limitations, and practical challenges associated with integrating autonomous LLM-based agents into real-world software development workflows. In particular, we identified 10 common challenges and four key areas for improvement. These findings can guide future research and practice toward improving the reliability, effectiveness, and practical adoption of LLM-based agent systems in software development. Future work will extend the evaluation to participants from multiple universities and industry organizations to provide a broader and more diverse assessment of the proposed system.

\section*{Acknowledgments}
This work is co-funded by the MAISA project (2025–2027) and the BF/Amalia-2023/SW project, both funded by Business Finland.






\printbibliography

@article{ouyang2022training,
  title={Training language models to follow instructions with human feedback},
  author={Ouyang, Long and Wu, Jeffrey and Jiang, Xu and Almeida, Diogo and Wainwright, Carroll and Mishkin, Pamela and Zhang, Chong and Agarwal, Sandhini and Slama, Katarina and Ray, Alex and others},
  journal={Advances in Neural Information Processing Systems},
  volume={35},
  pages={27730--27744},
  year={2022}
}

@inproceedings{feng2023investigating,
  title={Investigating Code Generation Performance of Chat-GPT with Crowdsourcing Social Data},
  author={Feng, Yunhe and Vanam, Sreecharan and Cherukupally, Manasa and Zheng, Weijian and Qiu, Meikang and Chen, Haihua},
  booktitle={Proceedings of the 47th IEEE Computer Software and Applications Conference},
  pages={1--10},
  year={2023}
}

@inproceedings{treude2023navigating,
  title={Navigating complexity in software engineering: A prototype for comparing gpt-n solutions},
  author={Treude, Christoph},
  booktitle={2023 IEEE/ACM 5th International Workshop on Bots in Software Engineering (BotSE)},
  pages={1--5},
  year={2023},
  organization={IEEE}
}

@article{li2022competition,
  title={Competition-level code generation with alphacode},
  author={Li, Yujia and Choi, David and Chung, Junyoung and Kushman, Nate and Schrittwieser, Julian and Leblond, R{\'e}mi and Eccles, Tom and Keeling, James and Gimeno, Felix and Dal Lago, Agustin and others},
  journal={Science},
  volume={378},
  number={6624},
  pages={1092--1097},
  year={2022},
  publisher={American Association for the Advancement of Science}
}

@article{chen2021evaluating,
  title={Evaluating large language models trained on code},
  author={Chen, Mark and Tworek, Jerry and Jun, Heewoo and Yuan, Qiming and Pinto, Henrique Ponde de Oliveira and Kaplan, Jared and Edwards, Harri and Burda, Yuri and Joseph, Nicholas and Brockman, Greg and others},
  journal={arXiv preprint arXiv:2107.03374},
  year={2021}
}

@inproceedings{fried2022incoder,
  author       = {Daniel Fried and
                  Armen Aghajanyan and
                  Jessy Lin and
                  Sida Wang and
                  Eric Wallace and
                  Freda Shi and
                  Ruiqi Zhong and
                  Scott Yih and
                  Luke Zettlemoyer and
                  Mike Lewis},
  title        = {InCoder: {A} Generative Model for Code Infilling and Synthesis},
  booktitle    = {The Eleventh International Conference on Learning Representations,
                  {ICLR} 2023, Kigali, Rwanda, May 1-5, 2023},
  publisher    = {OpenReview.net},
  year         = {2023},
  bibsource    = {dblp computer science bibliography, https://dblp.org}
}

@inproceedings{qian2023communicative,
   title={Chatdev: Communicative agents for software development},
  author={Qian, Chen and Liu, Wei and Liu, Hongzhang and Chen, Nuo and Dang, Yufan and Li, Jiahao and Yang, Cheng and Chen, Weize and Su, Yusheng and Cong, Xin and others},
  booktitle={Proceedings of the 62nd annual meeting of the association for computational linguistics (volume 1: Long papers)},
  pages={15174--15186},
  year={2024}
}

@inproceedings{zheng2023codegeex,
  title={Codegeex: A pre-trained model for code generation with multilingual benchmarking on humaneval-x},
  author={Zheng, Qinkai and Xia, Xiao and Zou, Xu and Dong, Yuxiao and Wang, Shan and Xue, Yufei and Shen, Lei and Wang, Zihan and Wang, Andi and Li, Yang and others},
  booktitle={Proceedings of the 29th ACM SIGKDD conference on knowledge discovery and data mining},
  pages={5673--5684},
  year={2023}
}

@article{hong2023metagpt,
  title={Metagpt: Meta programming for multi-agent collaborative framework},
  author={Hong, Sirui and Zheng, Xiawu and Chen, Jonathan and Cheng, Yuheng and Wang, Jinlin and Zhang, Ceyao and Wang, Zili and Yau, Steven Ka Shing and Lin, Zijuan and Zhou, Liyang and others},
  journal={arXiv preprint arXiv:2308.00352},
  year={2023}
}

@article{runeson2009guidelines,
  title={Guidelines for conducting and reporting case study research in software engineering},
  author={Runeson, Per and H{\"o}st, Martin},
  journal={Empirical Software Engineering},
  volume={14},
  pages={131--164},
  year={2009},
  publisher={Springer}
}

@article{yadav2024boldly,
  title={Boldly Going Where No Benchmark Has Gone Before: Exposing Bias and Shortcomings in Code Generation Evaluation},
  author={Yadav, Ankit and Singh, Mayank},
  journal={arXiv preprint arXiv:2401.03855},
  year={2024}
}

@article{dai2024mhpp,
  title={MHPP: Exploring the Capabilities and Limitations of Language Models Beyond Basic Code Generation},
  author={Dai, Jianbo and Lu, Jianqiao and Feng, Yunlong and Ruan, Rongju and Cheng, Ming and Tan, Haochen and Guo, Zhijiang},
  journal={arXiv preprint arXiv:2405.11430},
  year={2024}
}

@article{blair2015reflexive,
  title={A reflexive exploration of two qualitative data coding techniques},
  author={Blair, Erik},
  journal={Journal of Methods and Measurement in the Social Sciences},
  volume={6},
  number={1},
  pages={14--29},
  year={2015}
}

@article{huang2023agentcoder,
  title={Agentcoder: Multi-agent-based code generation with iterative testing and optimisation},
  author={Huang, Dong and Zhang, Jie M and Luck, Michael and Bu, Qingwen and Qing, Yuhao and Cui, Heming},
  journal={arXiv preprint arXiv:2312.13010},
  year={2023}
}

@article{yang2024swe,
  title={Swe-agent: Agent-computer interfaces enable automated software engineering},
  author={Yang, John and Jimenez, Carlos E and Wettig, Alexander and Lieret, Kilian and Yao, Shunyu and Narasimhan, Karthik and Press, Ofir},
  journal={Advances in Neural Information Processing Systems},
  volume={37},
  pages={50528--50652},
  year={2024}
}

@inproceedings{maleki2024ai,
  title={AI hallucinations: a misnomer worth clarifying},
  author={Maleki, Negar and Padmanabhan, Balaji and Dutta, Kaushik},
  booktitle={2024 IEEE conference on artificial intelligence (CAI)},
  pages={133--138},
  year={2024},
  organization={IEEE}
}

@inproceedings{islam2024mapcoder,
  title={Mapcoder: Multi-agent code generation for competitive problem solving},
  author={Islam, Md Ashraful and Ali, Mohammed Eunus and Parvez, Md Rizwan},
  booktitle={Proceedings of the 62nd Annual Meeting of the Association for Computational Linguistics (Volume 1: Long Papers)},
  pages={4912--4944},
  year={2024}
}

@article{wohlin2006empirical,
  title={Empirical research methods in web and software engineering},
  author={Wohlin, Claes and H{\"o}st, Martin and Henningsson, Kennet},
  journal={Web Engineering},
  pages={409--430},
  year={2006},
  publisher={Springer}
}

@article{bosch2015speed,
  title={Speed, data, and ecosystems: the future of software engineering},
  author={Bosch, Jan},
  journal={IEEE Software},
  volume={33},
  number={1},
  pages={82--88},
  year={2015},
  publisher={IEEE}
}

@article{liu2024large,
  title={Large language model-based agents for software engineering: A survey},
  author={Liu, Junwei and Wang, Kaixin and Chen, Yixuan and Peng, Xin and Chen, Zhenpeng and Zhang, Lingming and Lou, Yiling},
  journal={ACM Transactions on Software Engineering and Methodology},
  year={2024},
  publisher={ACM New York, NY}
}

@article{jin2024llms,
  title={From LLMs to LLM-based agents for software engineering: A survey of current, challenges and future},
  author={Jin, Haolin and Huang, Linghan and Cai, Haipeng and Yan, Jun and Li, Bo and Chen, Huaming},
  journal={arXiv preprint arXiv:2408.02479},
  year={2024}
}

@article{levy2021understanding,
  title={Understanding large-scale software systems--structure and flows},
  author={Levy, Omer and Feitelson, Dror G},
  journal={Empirical Software Engineering},
  volume={26},
  number={3},
  pages={48},
  year={2021},
  publisher={Springer}
}

@inproceedings{mohammadi2025evaluation,
  title={Evaluation and benchmarking of llm agents: A survey},
  author={Mohammadi, Mahmoud and Li, Yipeng and Lo, Jane and Yip, Wendy},
  booktitle={Proceedings of the 31st ACM SIGKDD Conference on Knowledge Discovery and Data Mining V. 2},
  pages={6129--6139},
  year={2025}
}

@article{hassan2025agentic,
  title={Agentic Software Engineering: Foundational Pillars and a Research Roadmap},
  author={Hassan, Ahmed E and Li, Hao and Lin, Dayi and Adams, Bram and Chen, Tse-Hsun and Kashiwa, Yutaro and Qiu, Dong},
  journal={arXiv preprint arXiv:2509.06216},
  year={2025}
}

@inproceedings{pashchenko2020qualitative,
  title={A qualitative study of dependency management and its security implications},
  author={Pashchenko, Ivan and Vu, Duc-Ly and Massacci, Fabio},
  booktitle={Proceedings of the 2020 ACM SIGSAC conference on computer and communications security},
  pages={1513--1531},
  year={2020}
}

@book{geraci1991ieee,
  title={IEEE standard computer dictionary: Compilation of IEEE standard computer glossaries},
  author={Geraci, Anne},
  year={1991},
  publisher={IEEE Press}
}

@inproceedings{paul2024benchmarks,
  title={Benchmarks and metrics for evaluations of code generation: A critical review},
  author={Paul, Debalina Ghosh and Zhu, Hong and Bayley, Ian},
  booktitle={2024 IEEE International Conference on Artificial Intelligence Testing (AITest)},
  pages={87--94},
  year={2024},
  organization={IEEE}
}

@misc{linaaker2015guidelines,
  author       = {Johan Lin{\aa}ker and Sardar Muhammad Sulaman and Rafael Maiani de Mello and Martin H{\"o}st},
  title        = {Guidelines for Conducting Surveys in Software Engineering},
  howpublished = {Department of Computer Science, Lund University},
  year         = {2015}
}

@misc{CodePori_Dataset,
  author = {Anonymous},
  title  = {CodePori: Large-Scale System for Autonomous Software Development Using Multi-Agent Technology},
  %year   = {2026},
  url    = {https://doi.org/10.5281/zenodo.17619419},
  note   = {Dataset}
}

@inproceedings{yao2022react,
  author       = {Shunyu Yao and
                  Jeffrey Zhao and
                  Dian Yu and
                  Nan Du and
                  Izhak Shafran and
                  Karthik R. Narasimhan and
                  Yuan Cao},
  title        = {ReAct: Synergizing Reasoning and Acting in Language Models},
  booktitle    = {The Eleventh International Conference on Learning Representations,
                  {ICLR} 2023, Kigali, Rwanda, May 1-5, 2023},
  publisher    = {OpenReview.net},
  year         = {2023},
  bibsource    = {dblp computer science bibliography, https://dblp.org}
}

@misc{CodePori_v02,
  author = {Anonymous},
  title  = {CodePori: Large-Scale Multi-Agent System for Autonomous Software Development},
  %year   = {2026},
  url    = {https://github.com/GPT-Laboratory/CodePori_v02},
  %note   = {Accessed: 2026}
}

@article{guggulothu2020code,
  title={Code smell detection using multi-label classification approach},
  author={Guggulothu, Thirupathi and Moiz, Salman Abdul},
  journal={Software Quality Journal},
  volume={28},
  number={3},
  pages={1063--1086},
  year={2020},
  publisher={Springer}
}

@article{host2000using,
  title={Using students as subjects—a comparative study of students and professionals in lead-time impact assessment},
  author={H{\"o}st, Martin and Regnell, Bj{\"o}rn and Wohlin, Claes},
  journal={Empirical Software Engineering},
  volume={5},
  number={3},
  pages={201--214},
  year={2000},
  publisher={Springer}
}

@article{xiong2025self,
  title={Self-organizing agent network for LLM-based workflow automation},
  author={Xiong, Yiming and Wang, Jian and Li, Bing and Zhu, Yuhan and Zhao, Yuqi},
  journal={arXiv preprint arXiv:2508.13732},
  year={2025}
}


\end{document}